\documentclass[12pt]{iopart}

\usepackage{graphicx}
\usepackage{booktabs}
\usepackage[separate-uncertainty=true]{siunitx}
\usepackage{xspace} 
\def\SiN{SiRN\xspace}
\def\SiO{SiO$_2$\xspace}

\begin{document}

\title{Capillary origami of micro-machined micro-objects: Bi-layer conductive hinges}

\author{A. Legrain $^1$, J.W. Berenschot $^1$, N.R. Tas $^1$ and L. Abelmann $^{1,2}$}
\address{$^1$ MESA+ Institute for Nanotechnology, University of Twente, Enschede, The Netherlands}
\address{$^2$ KIST Europe, Saarbrücken, Germany}

\ead{a.b.h.legrain@utwente.nl}

\date{\today}

\date{\today}

\begin{abstract}
Recently, we demonstrated controllable 3D self-folding by means of capillary forces of silicon-nitride micro-objects made of rigid plates connected to each other by flexible hinges~\cite{Legrain2014}. In this paper, we introduce platinum electrodes running from the substrate to the plates over these bendable hinges. The fabrication yield is as high as \SI{77(2)}{\percent} for hinges with a length less than \SI{75}{\um}. The yield reduces to \SI{18(2)}{\percent} when the length increases above \SI{100}{\um}. Most of the failures in conductivity are due to degradation of the platinum/chromium layer stack during the final plasma cleaning step. The bi-layer hinges survive the capillary folding process, even for extremely small bending radii of \SI{5}{\um}, nor does the bending have any impact on the conductivity. Stress in the different layers deforms the hinges, which does not affect the conductivity. Once assembled, the conductive hinges can withstand a current density of \SI{1.6(04)d6}{A/cm^2}. This introduction of conductive electrodes to elastocapillary self-folded silicon-based micro-objects extends the range of their possible applications by allowing an electronic functionality of the folded parts.
\end{abstract}

\maketitle
 
\section{Introduction}

Self-folding broadly refers to the self-assembly of interconnected parts that fold themselves into predefined shapes without the need of active human control. The assembly is either triggered by external stimuli (e.g, pH or temperature variation), enabled by external forces such as magnetic forces, or by internal forces such as pre-stressed layers~\cite{Leong2010,Shenoy2012,Ionov2013}. Three-dimensional assembly of micro/nano objects is possible using self-folding, where inherently two-dimensional micro-fabrication techniques have been shown to be inadequate~\cite{Madou1997}.  
 
\textit{Capillary origami} designates the self-folding of flexible elastic material using surface tension as the enabling force~\cite{Syms2003,Py2007}. Capillary origami is a particularly interesting tool for micro-fabrication, since at small scales, surface forces dominate over bulk forces such as gravity~\cite{Boncheva2003,Roman2010}. This technique was first employed to assemble silicon-based micro-objects by Syms~\emph{et al.}, who used melting solder to assemble hingeless silicon objects with integrated metal pads in flaps~\cite{Syms1993,Syms1995}. The scope of such solder assembly was then extended by Gracias~\emph{et al.}, who demonstrated the folding of complex structures (cubes, pyramids...) of micrometer and nanometer sizes fabricated using standard lithography  and deposition techniques~\cite{Gracias2002,Leong2007,Cho2009a,Cho2010a}. Applications of such structures range from three-dimensional micro-opto-electro-mechanical systems (MOEMS)~\cite{Syms2000,Linderman2002,Hong2006c} to RF nano-antennae~\cite{Park2011}.
 
We have used capillary origami to fold silicon nitride micro-objects. Unlike in solder assembly, the structures are necessarily hinged and their folding relies on the deformation of thin flexible silicon nitride plates, therefore the method can also be called elastocapillary folding. The folding is driven by the surface tension of water, which is simply manually deposited~\cite{vanHonschoten2010,vanHonschoten2011} or brought to the origami pattern through a tube at its center~\cite{Legrain2014}. The final shape  of the object is predefined by the patterning of rigid silicon nitride plates that form the body of the 3D object,  linked to each other by hinges. Objects remain assembled due to strong stiction between the flaps~\cite{vanHonschoten2010,Legrain2014} or by designing complex stop-programmable hinges~\cite{vanHonschoten2011}.

\begin{figure}
\centering
\includegraphics[width=.8\linewidth]{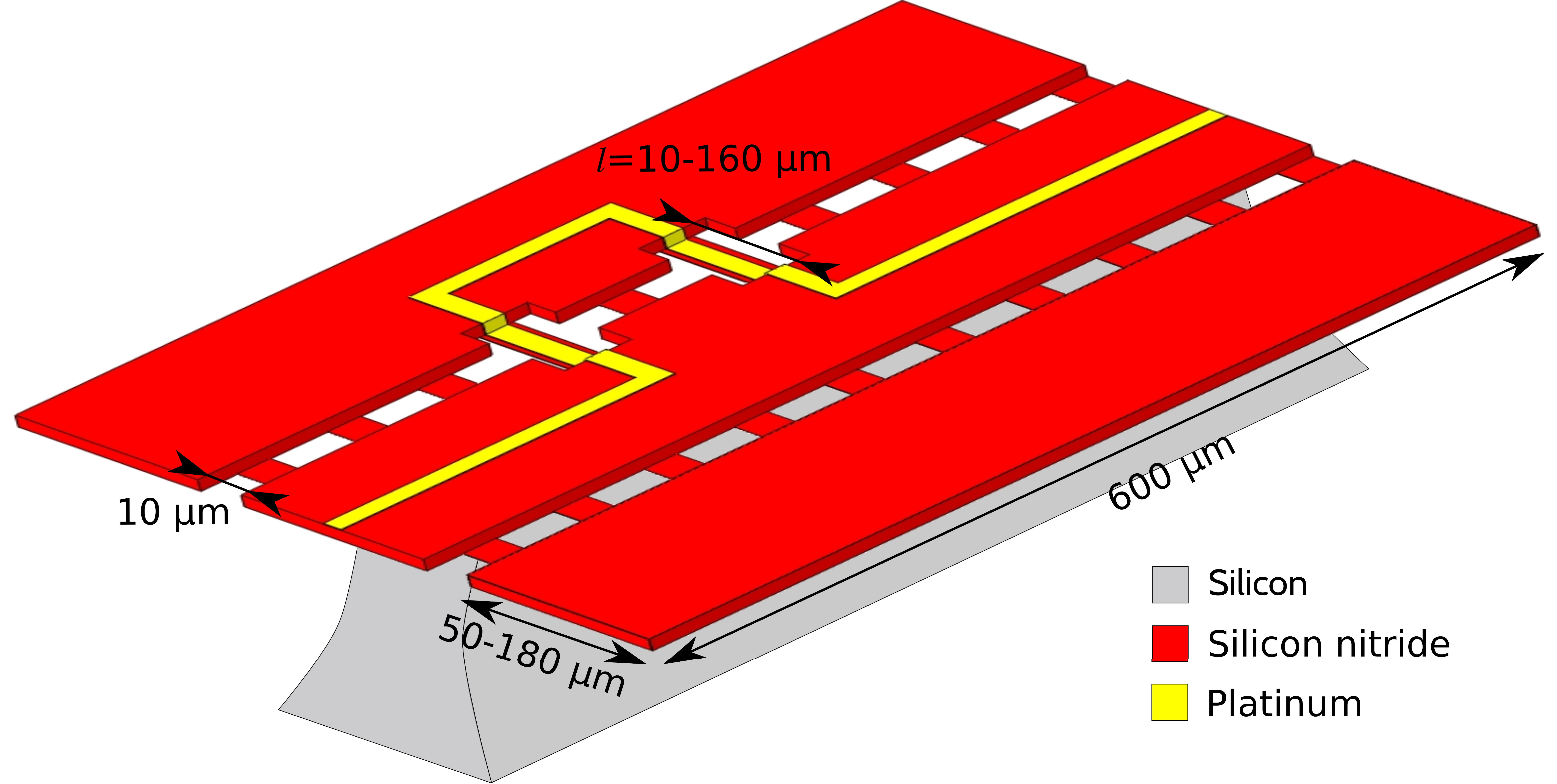}
\caption{Concept image of a typical test structure. Two free hanging silicon nitride flaps are connected to the central part by means of thin flexible silicon nitride hinges. The central part is fixed, resting on a silicon pillar. In this example, two decoupled bi-layer hinges run along the center of the structure. The length $l$ of the metallized junctures can be tuned, while the plain \SiN hinges are always 10~\si{\um} long. The metal parts run to contact pads on the outside of the structures.}
\label{fig:Principle}
\end{figure}

The applications of this self-folding technique are limited to situations where a passive mechanical structure is required that extends far above the wafer surface. To extend the range of applications, it would be extremely useful if electrical connection can be made to the moving parts of the folding structure. Electrical conductance over hinges has been demonstrated in experiments where folding is achieved by magnetic lifting~\cite{Chen2003,Chang2009} or stress gradients~\cite{Iker2006,Raskin2014}. In this research we have combined electrical connectivity with folding by capillary forces. Figure~\ref{fig:Principle} shows the type of structures used to prove the feasibility of conductive hinges.  Structures based on our previous publications~\cite{vanHonschoten2010,Legrain2014} are extended with platinum wires that run from the substrate towards the flap via bendable hinges. After elastocapillary folding, a three-dimensional triangular prism structure (`Toblerone') is realized,  which has electrical wiring on its movable parts.

In the following we will explain the fabrication process and demonstrate that conductivity can be preserved under folding. We will focus on the yield of the process and discuss various causes of failure during fabrication.

\section{Experimental section}

\subsection{Fabrication}
\label{part:fab}

\begin{figure}
\centering
\includegraphics[width=1\linewidth]{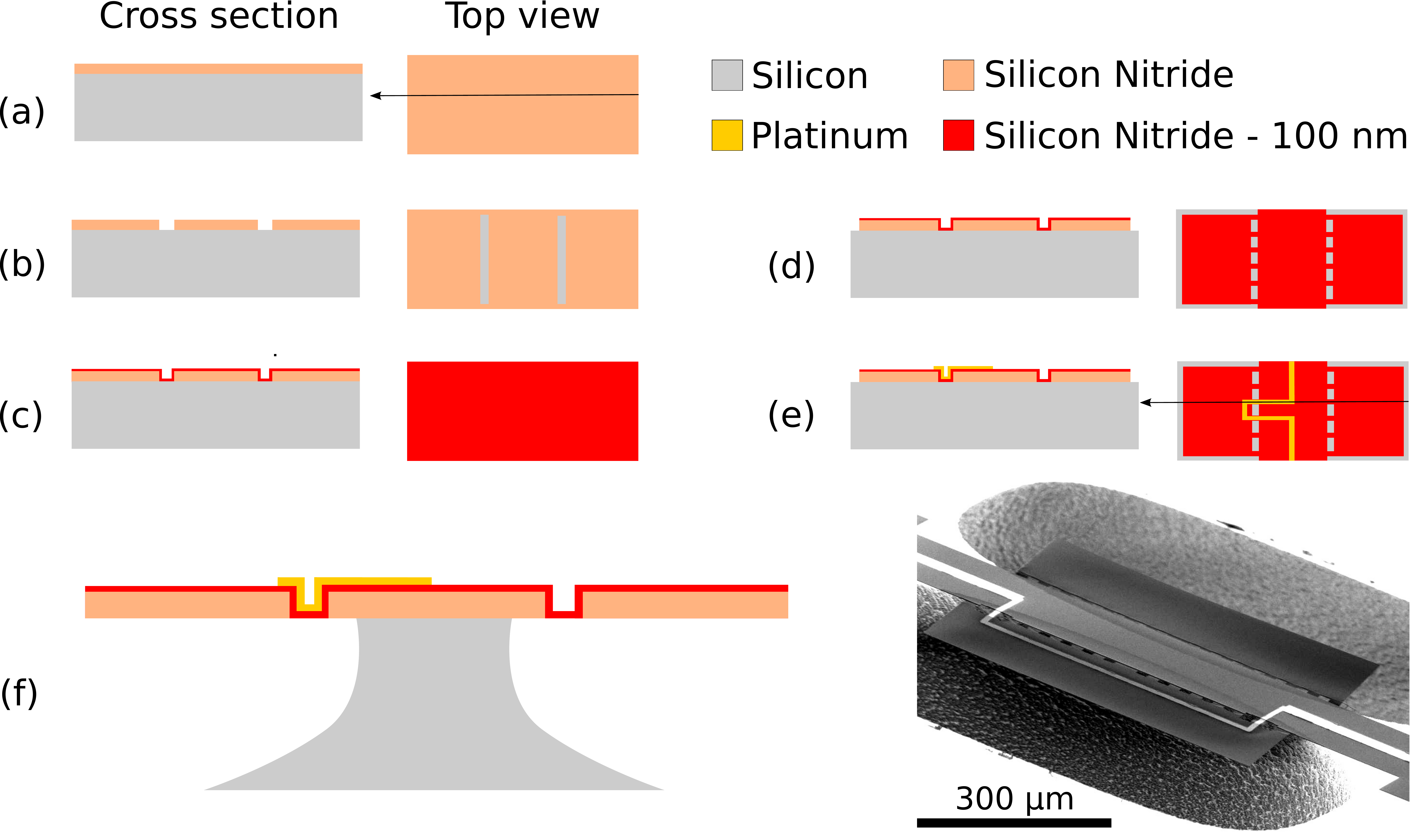}
\caption{Fabrication steps for silicon nitride (\SiN) structures with integrated conductive hinges capable of being folded out of the plane by capillary forces. (a): Deposition of \SiN by LPCVD. (b): First lithographic step; definition of the hinges. \SiN is either dry or wet etched. (c): Second deposition of \SiN . (d): Overall definition of the structures by a lithography step and subsequent dry etching. (e): Sputtering of metal followed by a standard lift-off procedure. (f): Release of the flexible objects by semi-isotropic etching of silicon. The SEM picture on the right hand side shows an example of a final structure.}
\label{fig:Process}
\end{figure}

Figure~\ref{fig:Process} shows the necessary steps to micro-machine test structures such as the one depicted in Figure~\ref{fig:Principle}. Fabrication starts with a conformal deposition by low pressure chemical vapor deposition (LPCVD) of a low stress thick silicon nitride (\SiN, Figure~\ref{fig:Process}-(a)). Three layers, 499, 797 and \SI{1083}{\nm}, were deposited on different samples to investigate the influence of the step height on the fabrication yield.

A first lithographic step follows to etch away the thick \SiN at the location of the future hinges. For this step, both dry and wet etching were used in order to study the impact on the folding of the shape of the transition between the hinges and the flaps.  Dry etching is performed after patterning only the photoresist, and  yields a straight profile in the thick \SiN layer. Wet etching requires first the deposition of a thin polysilicon layer ($\simeq \SI{50}{\nm}$). The polysilicon layer is patterned using dry etching, and subsequently used as a mask to etch \SiN in isotropic wet etchant, either hydrogen fluoride (HF) or phosphoric acid (H$_3$PO$_4$). The resulting shape is a smooth circular transition, inherent to isotropic etching. Once silicon is reached, the thin polysilicon layer can be stripped in tetramethylammonium hydroxide (TMAH). 

The next step is the deposition of a thin flexible layer of \SiN by LPCVD, Figure~\ref{fig:Process}-(c). A  thickness of \SI{100}{\nm} has proven to provide both flexibility and solidity to the hinges~\cite{vanHonschoten2010,Legrain2014}. 

The second lithography step follows to define the geometry of the patterns, Figure~\ref{fig:Process}-(d). In this design, combinations of different flap widths (50--\SI{180}{\um}),  leading to different final folding angles,  with different bi-layer hinge lengths (10--\SI{160}{\um}) were tried out. The location of the conductive hinges was also varied, and reference structures (structures without \SiN below metal or with no metal) were included.

Next, metal is deposited and patterned on top of the \SiN by a standard lift-off procedure, Figure~\ref{fig:Process}-(e).  Chromium ($\simeq \SI{10}{\nm}$) is used as an adhesive layer. Two different layers of platinum, \SI{75}{\nm} and \SI{150}{\nm}, were sputtered on different samples at a pressure of $\SI{6.6d-6}{\bar}$. The thicknesses of the final metal layers were checked using a mechanical surface profiler. Lift-off was achieved by first coating and patterning a \SI{3.5}{\um} thick photoresist with no post-development baking, to keep the profile of the resist straight. After metal deposition, the wafers were placed in an ultrasonic bath and immersed in sequence in acetone and in isopropanol.

Finally, a last lithography step is performed to protect the \SiN objects during their release, Figure~\ref{fig:Process}-(f). Under-etching of silicon is performed in semi-isotropic sulfur hexafluoride (SF$_6$) gas etchant. Since all the structures should be released simultaneously, the mask openings are designed to be of equal size and at the same distance from the final etch location. 

Moreover, it should be noted that the etching selectivity between silicon and \SiN during this step is around 1000. The initial thickness of the \SiN layer from which the hinges are patterned  (Figure~\ref{fig:Process}-(c)) was therefore chosen accordingly thicker ($\approx \SI{20}{\nm}$), to compensate for thinning during the silicon etch step.

Finally, the photoresist is stripped in oxygen plasma for two hours once the objects are released from the substrate.

\subsection{Residual stress measurements}
\label{Stress_measurements}

The stress in the deposited thin layers was measured using the wafer curvature method. The curvature of a dummy wafer is measured before and after deposition of the thin film by means of a surface profiler. Stress is then calculated using the Stoney equation~\cite{Stoney1909}. In the case of \SiN, for which material is deposited on both sides of wafers by LPCVD, one side of the wafer is stripped before measuring the radius of curvature. The thickness of the substrates is measured using a dedicated tool (Heidenhain measuring station) while the thickness of the \SiN and sputtered metals are checked by ellipsometry and mechanical surface profiler, respectively. The values found in the literature are used for Young's moduli and Poisson's ratios.

\subsection{Experimental setup}

\begin{figure}
\centering
\includegraphics[width=.8\linewidth]{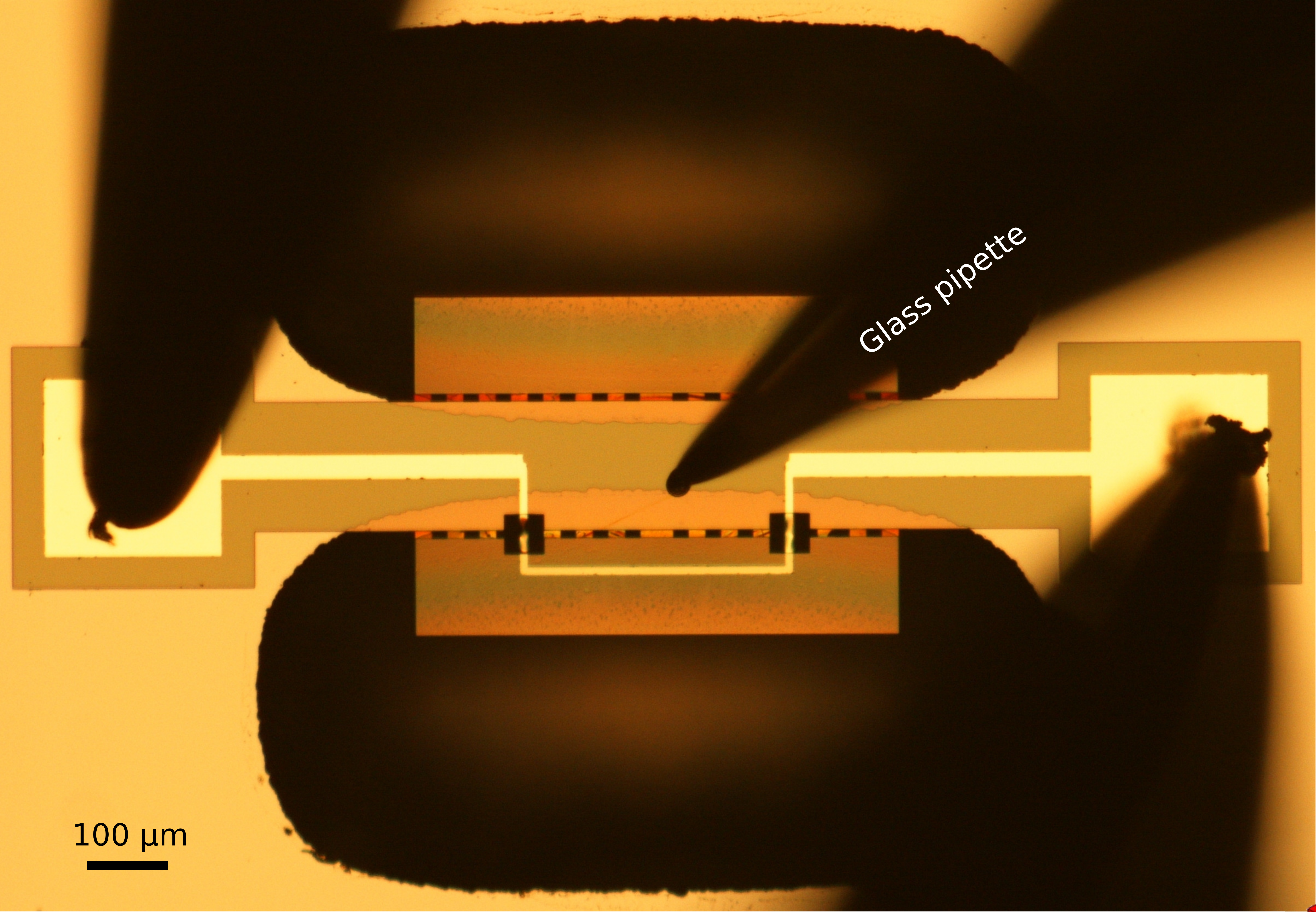}
\caption{Photograph of experimental setup, top view. The foldable object is placed at the center. A glass syringe is positioned on top of the structure by means of a translation table, and is used to apply water for the folding process. Metal probes are in contact with metal pads on both sides.}
\label{fig:Setup}
\end{figure}

Self-folding experiments and the electrical characterization are carried out in the same setup, shown in Figure~\ref{fig:Setup}. Resistivity of the structures is measured by a multimeter \emph{in situ} during assembly by means of two metal probes placed on metal contact pads on both sides of the structures. For breakdown measurements, a voltage source was used to force a current through the bi-layer hinges.

A glass micro-pipette of \SI{10}{\um} diameter is positioned on top of the micro-origami pattern by an accurate \emph{x-y} translation table. A low volume syringe ($\sim \SI{10}{\uL}$), filled with ultra pure water, is used to propel a drop out of the micro-pipette. A hydrophobic  polytetrafluoroethylene  (teflon) coating is applied on the micro-pipette to avoid wetting of the outside of the glass pipette. Two cameras on top and on the side are used to monitor the folding process. Depending on the volume of liquid deposited, folding takes 10 to \SI{120}{\second}.

\section{Results and discussion}

\subsection{Fabrication yield}

As described in Part~\ref{part:fab}, various combinations of metal thicknesses  and step heights have been fabricated. Moreover, the transition between the hinges and the flaps was machined by dry etching, resulting in a sharp step,  as well as wet etching, resulting in a smooth circular transition.

None of the aforementioned parameters have any significant influence on the fabrication yield. The only feature that proved to have a statistically significant impact on the yield is the length of the composite hinges. Table~\ref{tab:yield} summarizes our results. Short hinges, with a length $l$ below \SI{75}{\um}, are the most robust, with a yield of \SI{77(2)}{\percent}. The variance of this value is calculated from the number of structures tested assuming a binomial distribution. The yield drastically drops for longer junctures, and less than one out of five structures are conductive when $l \ge 100~\si{\um}$. The reasons will be made clear in Section~\ref{Hinges_deformations}. Short hinges are therefore the best option for the design of conductive bi-layer junctures.

\begin{table}
\centering
\begin{tabular}{@{}rrrcc@{}}
\toprule
\multicolumn{3}{c}{Hinge length} & Conductivity yield & $N$\\ 
\midrule 
10 &$\le$  $l$ $\le$ &\SI{75}{\um}   & \SI{77(2)}{\percent} & 541  \\ 
75 &$<$    $l$ $\le$ &\SI{100}{\um} & \SI{26(4)}{\percent} & 129  \\ 
              & $l$  $>$  &\SI{100}{\um}  & \SI{18(2)}{\percent} & 391  \\ 
\bottomrule
\end{tabular}
\caption{Fabrication yield of conductive hinges as a function of hinge length, before folding. A correlation between length and conductivity is obvious. Statistics made with $N$ structures from 7 wafers, which were fabricated with different combinations of thicknesses for the thick \SiN flaps (499, 797 or \SI{1083}{nm}), thicknesses of platinum (75 or \SI{150}{nm}) and etching method for hinges (dry or wet, Figure~\ref{fig:Process}-(b)). The latter parameters had no observable significant impact on the yield.}
\label{tab:yield} 
\end{table}

\subsection{Folding and electrical characterization of conductive hinges}
\label{sec:elec_degrad}

\begin{figure}
\centering
\includegraphics[width=.8\linewidth]{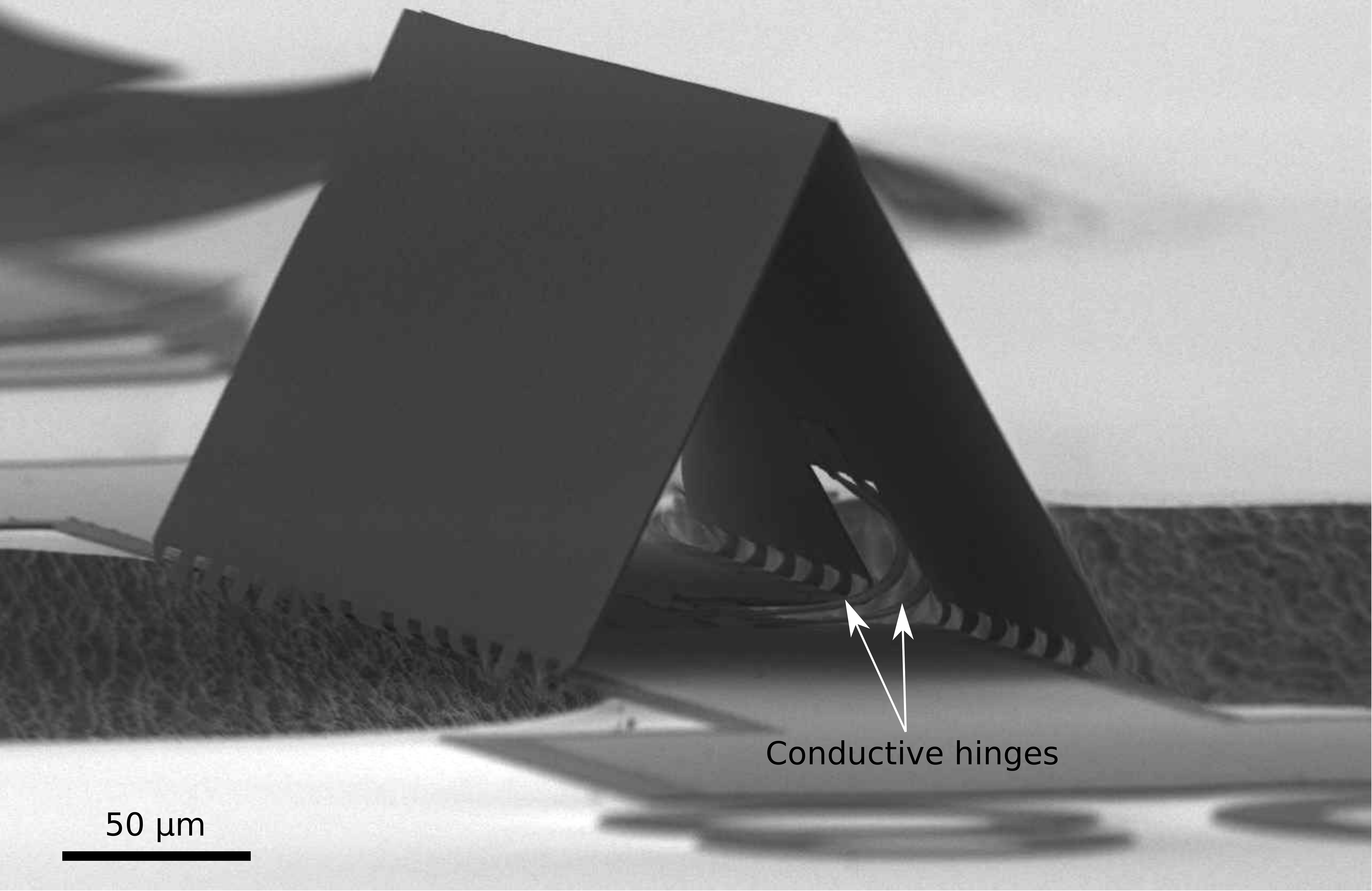}
\caption{Folded micro-object with conductive bi-layer hinges that allow for electrical connection to the folded flaps. The structure is \SI{600}{\um} long and the flaps are \SI{80}{\um} wide. The plain \SiN hinges are \SI{10}{\um} long and \SI{30}{\um} wide, while the two bi-layer hinges are \SI{75}{\um} long.}
\label{fig:Final_structures}
\end{figure}

Fifty structures with conductive bi-layer hinges were successfully folded and monitored using the setup presented in Figure~\ref{fig:Setup}. No breaking or change in the conductivity of the hinges was observed. Figure~\ref{fig:Final_structures} shows an example of a folded micro-object, with the two conductive hinges inside the right-hand side flap. Neither the folding angle nor the length of the bi-layer hinges have any impact on the folding. As shown in Figure~\ref{fig:short}, the conductive hinges do not break even when they are rotated by \SI{180}{\degree}. Picture (b) shows \SI{10}{\um} bi-layer hinges that are fully rotated and yet do not break, therefore resulting in a radius of curvature $R\sim \SI{5}{\um}$. These are the shortest that can be designed since they are as long as the plain \SiN hinges. We can therefore conclude that the folding of metallized hinges does not yield a stress higher than their failure stress, even in extreme conditions. Such extremely small radii can be found in recent publications on flexible electronics. Using ultra-thin substrates ($<\SI{2}{\um}$), fully printed organic TFTs could be folded to very small bending radii of about $R\sim \SI{5}{\um}$~\cite{Fukuda2014}. Bendable organic solar cells   with $R\sim \SI{10}{\um}$~\cite{Kaltenbrunner2012},  as well as polymer-based LEDs with $R< \SI{10}{\um}$~\cite{White2013}, have also been reported.

 The resistivity between the two metal probes was measured to be on the order of \SI{110}{\ohm} and \SI{80}{\ohm} for a platinum thickness of \SI{75}{\nm} and \SI{150}{\nm}, respectively. The value varies by \SI{25}{\percent}, depending on the total length of the metal and the exact landing position of the probes. These values are higher than the theoretical values of respectively \SI{64(18)}{\ohm} and \SI{32(5)}{\ohm}, calculated from the bulk value of the resistivity of platinum, $\rho_\text{b} =\SI{1.06d-7}{\ohm\metre}$. The difference can be explained by additional contact resistances as well as the fact that  the resistivity of  platinum thin films is known to be higher than the bulk value. A factor 5 is reported for \SI{20}{\nm} platinum thick films~\cite{Avrekh2000,Salvadori2004}. It seems therefore likely that the resistivity is larger than the bulk value in our situation, especially for a platinum thickness of \SI{75}{\nm}.

The current passing through the hinges was measured while applying an increasing potential difference at their terminals. The resistance increases at higher voltages, as is illustrated by a linear fit through the points below \SI{1}{V}.  It is likely that this increase is caused by an increase in temperature. The measured resistance, however, includes the series resistances and therefore cannot be used to estimate the temperature of the metal layers.

\begin{figure}
\centering
\includegraphics[width=.8\linewidth]{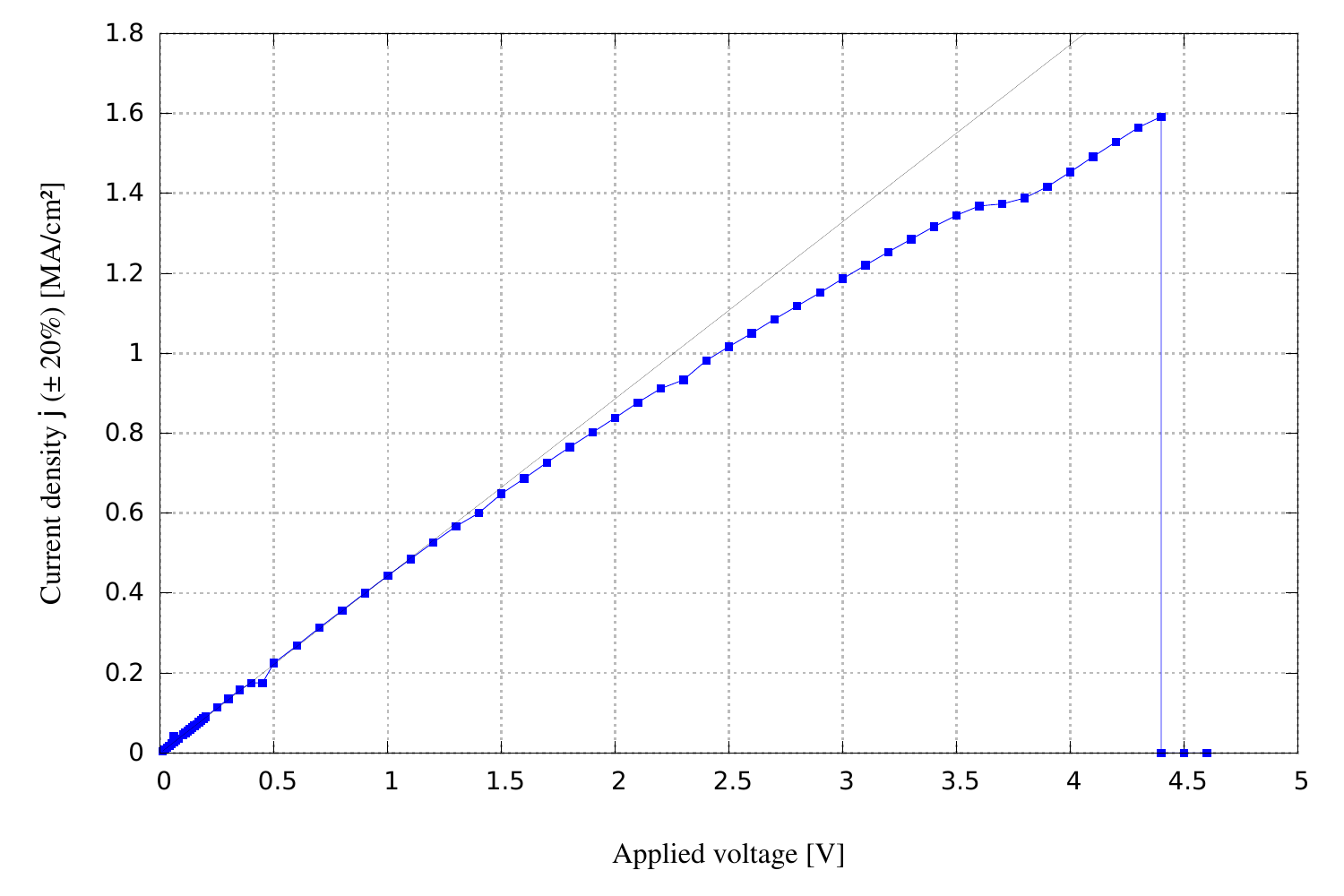}
\caption{Current--voltage characteristic of the bi-layer conductive hinges. The conductivity breaks down at $j=\SI{1.6(04)d6}{A/cm^2}$. The solid black line is a linear fit through the points below \SI{1}{V}.}
\label{fig:Graph_V_j}
\end{figure}

The bi-layer hinges lost conductivity during the experiments when a current density of $j=\SI{1.6(04)d6}{A/cm^2}$ was passing through them, as shown in Figure~\ref{fig:Graph_V_j}. This value of current density is close to the reported values in the literature at which electromigration has been observed~\cite{Kumar2011,Rudneva2013}. However, the failure could also be attributed to the high temperature induced by the high current density, which is linked to the degradation of platinum films~\cite{Tiggelaar2009}. In this case the dominant mechanism is agglomeration, a surface-diffusion driven capillary process, whose effects are worsened by a mismatch of the temperature coefficients between the different layers~\cite{Firebaugh1998}.  It is possible to limit the impact of high temperatures on the hinges, through annealing or depositing platinum without an adhesion layer~\cite{alepee2001}.

\subsection{Hinge deformation}
\label{Hinges_deformations}

\begin{figure}
\centering
\includegraphics[width=1\linewidth]{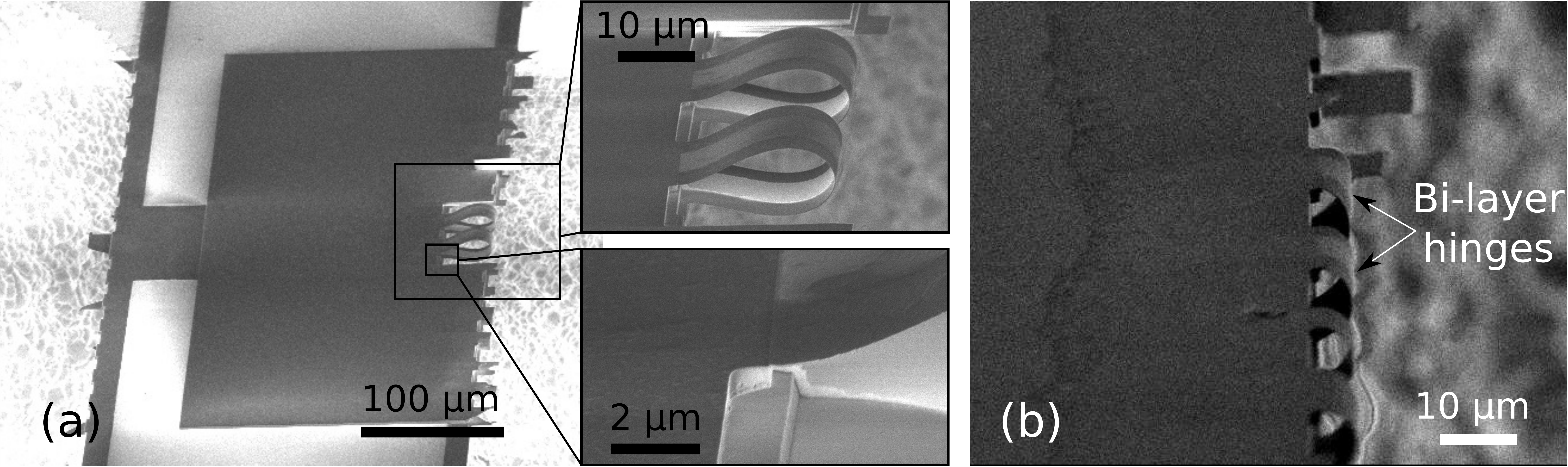}
\caption{(a): Extreme folding of \SI{50}{\um} long hinges over an angle of \SI{180}{\degree}. Top inset: a zoom on the formed loop, no deformation of the bi-layer junctures can be observed. Bottom inset: a zoom on the transition between the flaps and the bi-layer junctures. The standard sputtering method yields a nice conformal step coverage. (b): Even the shortest hinges of \SI{10}{\um} length  sustain a \SI{180}{\degree} rotation.}
\label{fig:short}
\end{figure}

\begin{figure}
\centering
\includegraphics[width=1\linewidth]{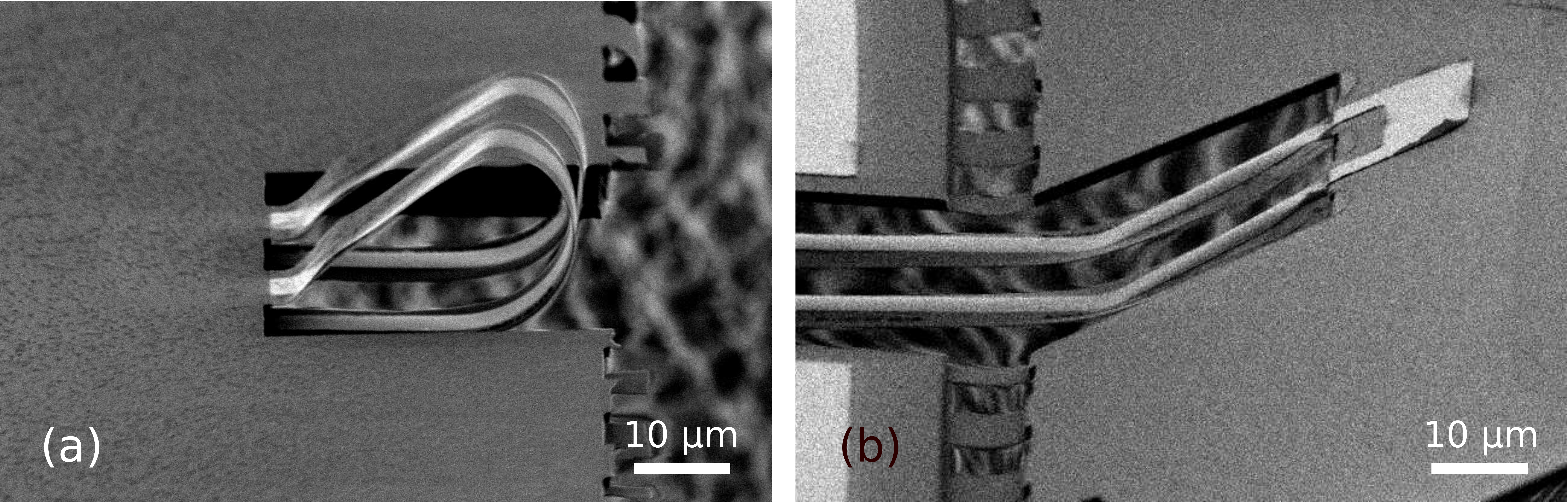}
\caption{Bi-layer hinges from two distinct structures. (a): $l=\SI{100}{\um}$, was folded by 180\si{\degree} using the setup shown in Figure~\ref{fig:Setup} (b): $l=\SI{105}{\um}$, was not folded. The flap is tilted out of plane (around \SI{30}{\degree}). Both structures are still conductive but show deformations due to stress mismatch between the different layers. In both cases only the parts located at the level of the flaps are deformed.}
\label{fig:medium}
\end{figure}

While short hinges show no deformation and are nicely curved after folding, see Figure~\ref{fig:short}, longer junctures are deformed, see Figure~\ref{fig:medium}. The deformation occurs only at the level of the flaps, with the hinge section connected to the fixed central part remaining intact. In photograph (b) we can see that the flap is tilted out of the plane, which suggests that the deformation in the bi-layer junctures is due to this pre-folding.  Despite the strong deformation, the hinges are still conductive. 

\begin{figure}
\centering
\includegraphics[width=1\linewidth]{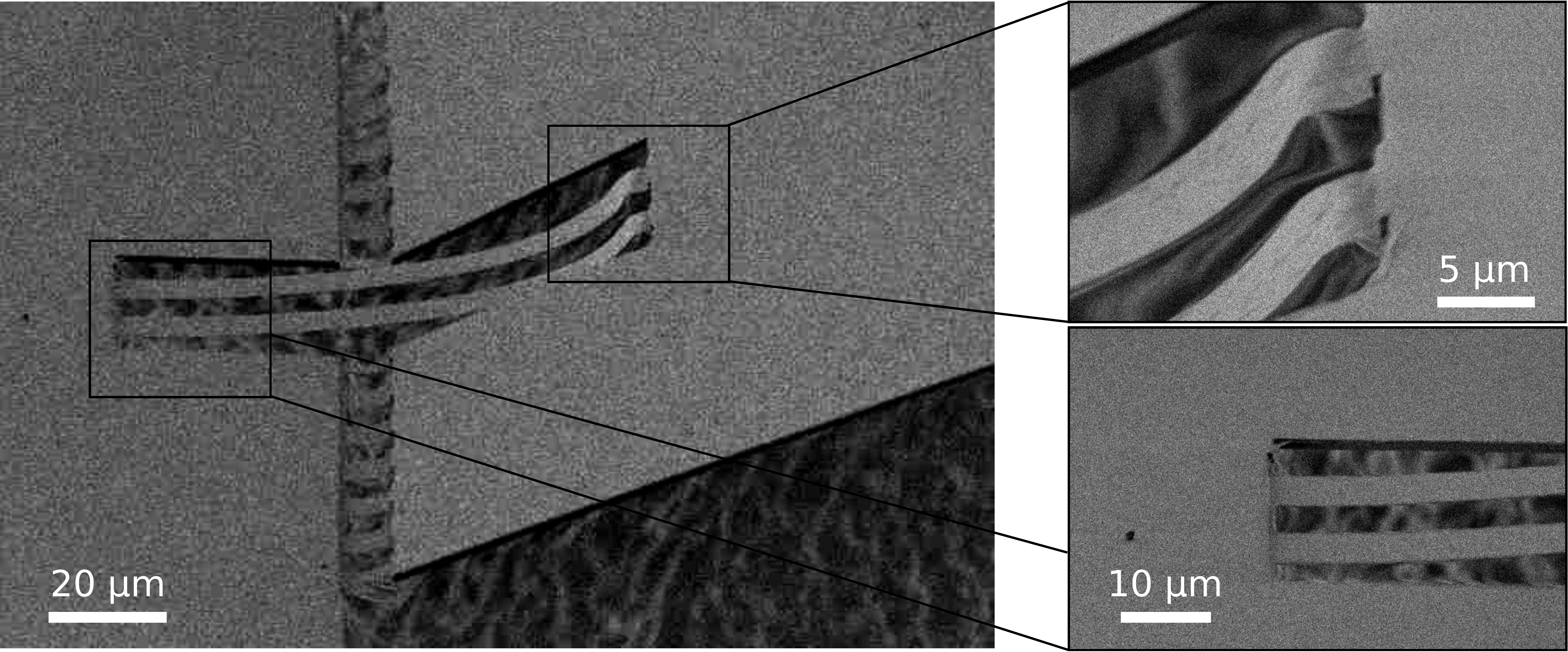}
\caption{Plain decoupled \SiN hinges. Although no metal is present on top of the junctures, the deformation already seen for bi-layer objects can be observed here. Most of the constraint seems to happen at the transition from the flap to the thin junctures.}
\label{fig:SiN_buckling}
\end{figure}

Figure~\ref{fig:SiN_buckling} shows similar hinges without a metal top layer. These plain decoupled \SiN hinges are deformed in a similar way as the bi-layer version of Figure~\ref{fig:medium}. As before, the section of the hinges connected to the central part remains intact. The other side, connected to the flap, remains parallel to the thick \SiN plane for a short distance before the curvature starts.

Why this deformation occurs is unclear. We can imagine several possible explanations for this effect, none of which seem adequate. The flaps are up to ten times thicker than the hinges, and the stress mismatch between the two \SiN layers could explain the plastic deformation. However, there is no evidence of a thickness dependency of the residual stress in the \SiN layers in the literature. \SiO layers between the two \SiN layers or under the \SiN formed as the wafers are loaded in the LPCVD equipment are also candidates. But such layers would be a few \SI{}{\nm} thick, and it is unlikely that they could be the cause of such large bending. We also observed that during the stripping of the photoresist in the oxygen plasma, the last step of fabrication, the temperature rises and the photoresist melts, which results in folding of the flaps. As will be discussed in the next section, we observed the degradation of metal during this same step. Such a degradation could be attributed to the diffusion of chromium,
which starts at \SI{300}{\degreeCelsius}. The stresses due to melting photoresist at rather high temperatures may cause permanent plastic deformation of the hinges. The correct explanation however is probably a combination of the previously described factors, coupled with the complex three-dimensional shape of the flap--hinges system.


Next to curvature along the hinges, buckling in the transverse direction is present in the case of bi-layer hinges, see Figure~\ref{fig:medium}. This buckling is not visible in the case of plain \SiN hinges, Figure~\ref{fig:SiN_buckling}. Since we suspect that this is caused by stress gradients, we measured the residual stress using a wafer curvature method, see Part~\ref{Stress_measurements}. The \SiN shows tensile stress, $\sigma_{r-SiN}= \SI{-169(25)}{\MPa}$, whereas the stress is compressive for platinum, $\sigma_{r-Pt} =\SI{+452(70)}{MPa}$. This significant mismatch causes a stress gradient along the bi-layer stack, which could explain the observed deformation.

\subsection{Failure mechanism}

\begin{figure}
\centering
\includegraphics[width=1\linewidth]{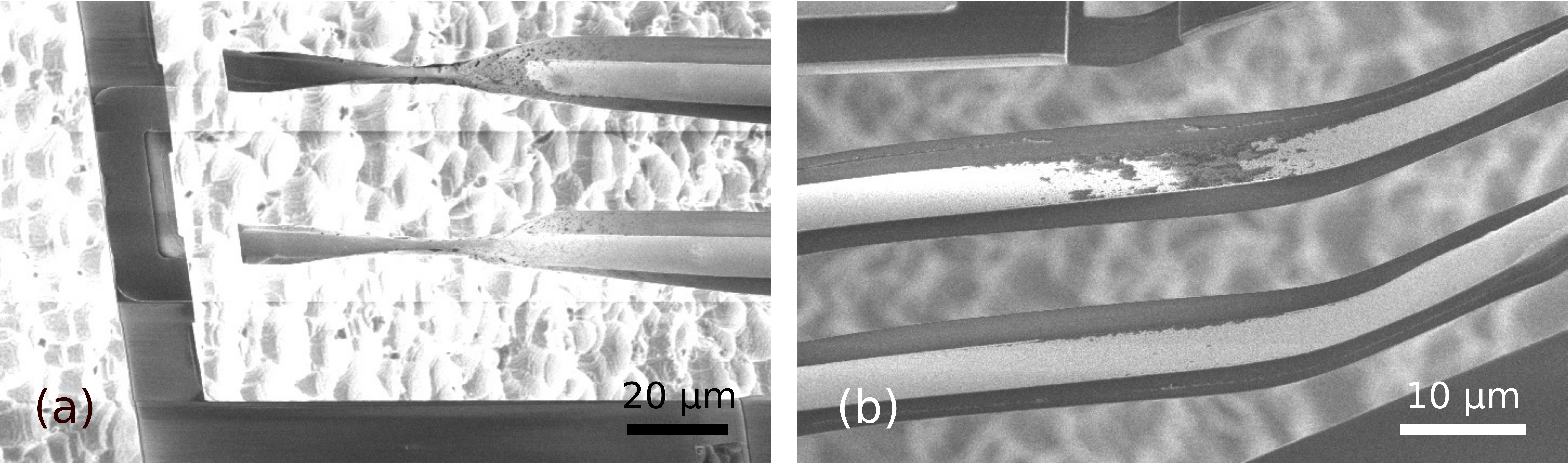}
\caption{Typical failures of long hinges that show no conductivity. (a): $l=160~\si{\um}$. For very long junctures, buckling and bending can cause the hinges to break at the transition with the flaps. (b): $l=125~\si{\um}$. Platinum burns at the middle of the hinges. }
\label{fig:long}
\end{figure}

Deformations caused by stress can cause the bi-layer hinges to break, as shown in Figure~\ref{fig:long}-(a). However, this is the case only for very long hinges, typically $l>\SI{150}{\um}$. Most of the long hinges lose conductivity because of a poor metal coverage following the final oxygen plasma cleaning step, as shown in Figure~\ref{fig:long}-(b). Why this deterioration of the platinum happens is unclear. Electrical discharge and heating during the two hour final oxygen plasma cleaning step seems to be the only plausible explanation.

\section{Conclusion}

We have successfully fabricated free-standing silicon-nitride plates with integrated platinum electrodes, which can be folded by elastocapillary forces into three-dimensional objects of several hundred \si{\um} in size. The electrodes are guided to the free-standing folding elements of the objects over two bi-layer silicon nitride/platinum hinges.

The resistance of the hinges does not change during the folding. A current density as high as $j=\SI{1.6(04)d6}{A/cm^2}$ was passed through the hinges before conductivity was lost. 

 None of the fifty bi-layer hinges, fabricated with different lengths, broke during folding. Bending radii as small as \SI{5}{\um} were achieved without mechanical failure or lost of conductivity.

If hinges break, it is during the fabrication process. Hinges with a length below \SI{75}{\um} were mostly conductive with a high yield of \SI{72(2)}{\percent}. The yield of the fabrication process drops to \SI{18(2)}{\percent} for bi-layer hinges with a length above \SI{100}{\um}. The yield is not influenced by the thickness of the connected flaps, the platinum layer, or the method used to etch the hinges' molds (wet or dry). 

 Short hinges with a length below \SI{50}{\um} are nicely curved and show no deformation or burning after folding.  Most of the mid-sized hinges, from 50 to \SI{75}{\um}, do not break, but do show visible deformations such as buckling and bending. Similar deformations are observed even when there is no metal on top of the hinges. Although made of the same material as the hinges, the connected flap is folded out of plane by around \SI{30}{\degree}.   

The failure mechanism during fabrication is either breaking or burning of the top platinum layer. In the case of long hinges with a length over \SI{100}{\um}, the stress mismatch in the layers causes the breaking of the hinges at their connection with the flaps in rare cases. The majority of failures is, however, due to burning of the platinum top layer which is observed after the final oxygen plasma cleaning step, leading to a loss of conductivity.

Elastocapillary folding is a powerful technique that allows the 3D assembly of silicon based objects of mm size, something which is known to be difficult to accomplish using inherently two-dimensional fabrication techniques. However, the self-folding method has lacked applications so far. We believe that these conductive hinges will extend the application scope of elastocapillary folded structures, with possible applications to out-of-plane sensing, high aspect ratio coils, or even 3D electronics.
\section*{Acknowledgements}
The authors would like to thank R.~G.~P.~Sanders for his valued help with the folding and electrical experiments, K.~Ma for his precious help with cleanroom work, and C.~M.~Bruinink  for his assistance with the residual stress measurements. This paper would have never been possible without J.~W.~van~Honschoten, who inspired this project.
\section*{References}


\end{document}